\documentclass[conference]{IEEEtran}
\IEEEoverridecommandlockouts
\usepackage{cite}
\usepackage{amsmath,amssymb,amsfonts}
\usepackage{algorithmic}
\usepackage{graphicx}
\usepackage{textcomp}
\usepackage{xcolor}
\usepackage{hyperref}
\def\BibTeX{{\rm B\kern-.05em{\sc i\kern-.025em b}\kern-.08em
    T\kern-.1667em\lower.7ex\hbox{E}\kern-.125emX}}
\begin{document}

\title{
A Flexible and Intelligent Framework for Remote Health Monitoring Dashboards
}

\author{\IEEEauthorblockN{Shayan Fazeli}
\IEEEauthorblockA{\textit{Computer Science Department} \\
\textit{UCLA}\\
Los Angeles, USA \\
shayan@cs.ucla.edu}
\and
% \IEEEauthorblockN{Sajad Darabi}
% \IEEEauthorblockA{\textit{Computer Science Department} \\
% \textit{UCLA}\\
% Los Angeles, USA \\
% shayan@cs.ucla.edu}
% \and
\IEEEauthorblockN{Majid Sarrafzadeh}
\IEEEauthorblockA{\textit{Computer Science Department} \\
\textit{UCLA}\\
Los Angeles, USA \\
majid@cs.ucla.edu}
}

\maketitle

\begin{abstract}
Developing and maintaining monitoring panels is undoubtedly the main task in the remote patient monitoring (RPM) systems. Due to the significant variations in desired functionalities, data sources, and objectives, designing an efficient dashboard that responds to the various needs in an RPM project is generally a cumbersome task to carry out. In this work, we present ViSierra, a framework for designing data monitoring dashboards in RPM projects. The abstractions and different components of this open-source project are explained, and examples are provided to support our claim concerning the effectiveness of this framework in preparing fast, efficient, and accurate monitoring platforms with minimal coding. These platforms will cover all the necessary aspects in a traditional RPM project and combine them with novel functionalities such as machine learning solutions and provide better data analysis instruments for the experts to track the information.
\end{abstract}

\begin{IEEEkeywords}
remote health monitoring, machine learning, flask, python, questionnaires, ehealth
\end{IEEEkeywords}

\section{Introduction}
The use of computerized methodologies in eHealth and medical research has been considerably growing during the past decade. Advancements in hardware and software technologies have rendered telemonitoring an immensely beneficial instrument for decreasing the expensive cost of in-patient monitoring both for the patient and on the side of health-care centers. In this work, we present a web-application design framework with a focus on remote patient monitoring (RPM) as its primary application and motivation. 

The recent literature indicates the growth of medical applications that include remote monitoring as their core component for improving the quality of patient care (e.g., self-care in chronic disease management \cite{bayliss2003descriptions}. Remote health monitoring has proven to be an extremely useful tool by causing significant improvements in the quality of care in different applications such as controlling and managing diabetes (Chase 2003), physical and mental health monitoring for veterans \cite{darkins2008care}, monitoring for infertility \cite{chausiaux2011pregnancy}, etc. Given the extensive domain of RPMs applications in health-care, different dashboards and portals on the receiving end (both for developers and clients) require various functionalities that generally make implementing such dashboards a cumbersome task to carry out.

In this article, we introduce ViSierra, a framework for remote monitoring dashboards by providing comprehensible and inclusive abstractions for various components that might be needed in such use cases. It also offers ways to further equip the main panel with machine learning and data analytics routines to shed more light on the underlying patterns in the incoming and stored data. This project is now open-source, and the codes and tutorials are available online\footnote{
    The source code, tutorial, and examples regarding this work are available at \href{https://github.com/shayanfazeli/visierra}{https://github.com/shayanfazeli/visierra}
}.

\section{Related Works}
Remote health monitoring is a widespread topic in eHealth research and has attracted numerous researchers to utilize it in their works. Retrieving and monitoring data coming from the body area sensor network resembles a well-understood use case of this approach. In \cite{kirbacs2012healthface} a web-based interface has been proposed for the health-care systems that are working based on body area sensor networks. Such networks are placed near, on, or within the human body and gather information on entities such as temperature, blood pressure, etc. This scheme requires medical practitioners at the health-care centers to be able to monitor the incoming data as well. The aforementioned web interface (a.k.a HealthFace) would assist them in performing this task. Their system was different than the other works presented in the literature in the sense that it was built using the MATLAB Builder NE, rendering it able to utilize a flexible environment to visualize the data and perform low-level data analysis on it. When it comes to medical applications, another consequential factor of providing web-based services is maintaining the integrity, confidentiality, and authenticity of our data. The HealthFace interface also prevents unauthorized access and requires user authentication to render any data observation and utilization accessible.

\begin{figure}
    \centering
    \includegraphics[width=0.5\textwidth]{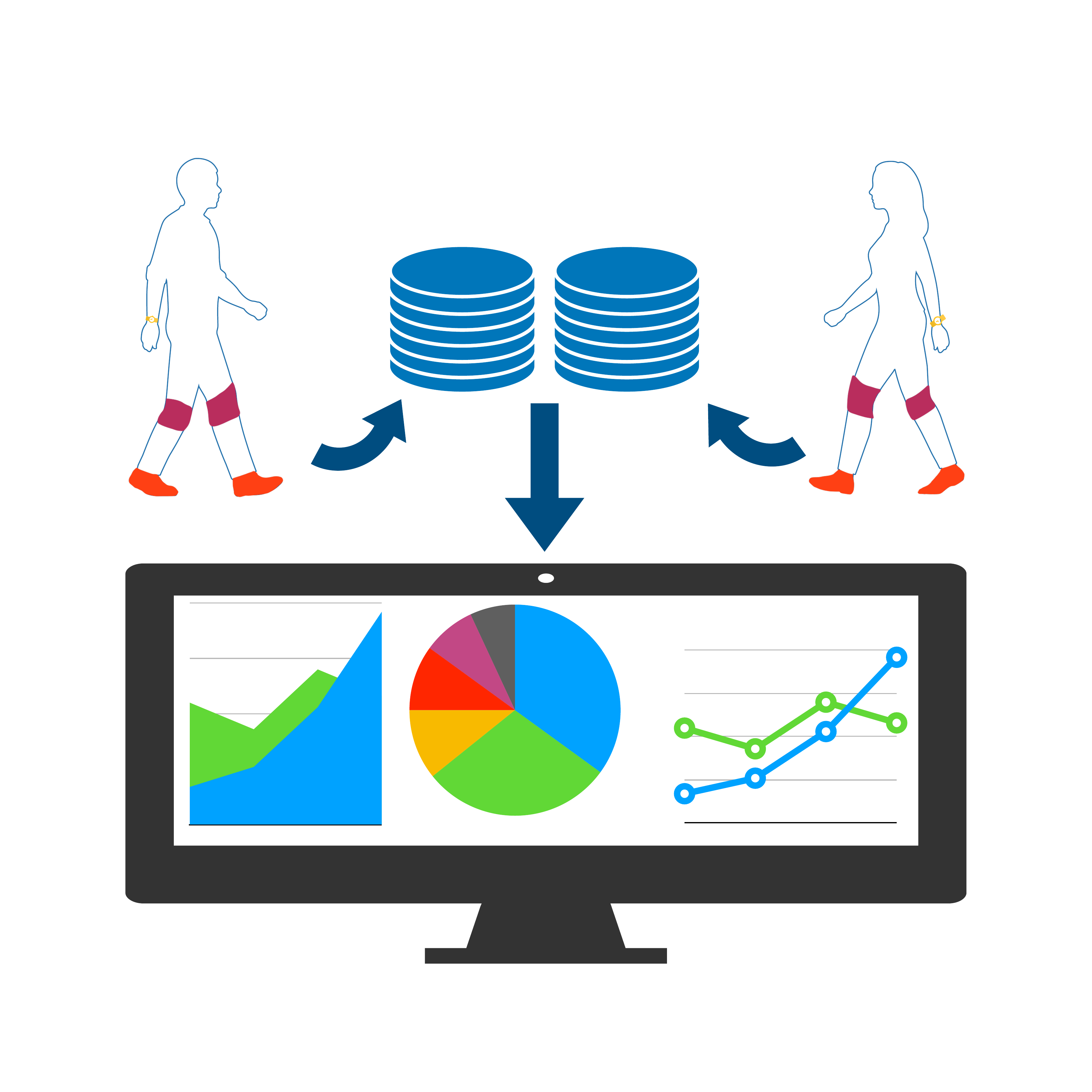}
    \caption{General diagram of a Remote Patient Monitoring system - Information is collected from subjects in the cohort. Our framework mainly has to do with the data analytics portals which ought to provide their audience with in-depth analyses over the stored data and depict informative visualizations to illustrate the statistical properties of data better.}
    \label{fig:overall pipeline}
\end{figure}

Mushcab et al. have proposed a survey on web-based systems' usage in remote monitoring as a means for self-managing type two diabetes \cite{mushcab2015web}. In their review, around four hundred publications were found focusing on diabetes-related studies and incorporating various methods of remote monitoring of the contributing factors. This number of research works indicate the prevalence of remote monitoring research in the eHealth domain.

As another domain of application for RPM, \cite{healy2010web} proposed an RPM system for monitoring the live electroencephalogram (EEG) feed in the critical care units. They used Adobe flash plugin to prepare a nearly-real-time updateable plot of the patients' EEG that was being transmitted through the world-wide-web and allow medical practitioners to review the results in a remote location. The motivation was the lack of expert availability at most of these units in an on-site manner, therefore, enabling experts to monitor and interpret the data using the proposed application remotely.

In monitoring heart-related diseases, WANDA is a well-known example of \cite{lan2012wanda}. WANDA includes a remote monitoring application for improving the quality of outpatient health-care for patients suffering from chronic heart failure. Another heart failure analysis dashboard was introduced in \cite{guidi2012heart} that was equipped with a computer decision support system to help the non-specialist personnel in making decisions regarding the severity of patients' conditions. Their system also included simple artificial intelligence components (e.g., SVMs) designed to perform patient classification. Another instance of RPM for cardiac patients is presented in \cite{kakria2015real}, further indicating the importance of such monitoring in the medical domain.

In \cite{sebestyen2010remote} authors focused on formalizing a framework for remote monitoring as it comprises the main component of almost all health-care applications, focusing on wearable sensors.

Our work differs from the current literature in the following aspects:

\begin{itemize}
    \item{
        We present a framework for designing web-based remote health monitoring applications by abstracting the major components that are needed in relevant use cases.
    }
    \item{
        The effectiveness of the proposed framework in real-world settings is discussed.
    }
    \item{
        The open-source implementation of the framework is made available to the unconstrained public use.
    }
    \item{
        Our system, being based on Python and Data-Driven Documents, provides great users with great freedom in customizing the given tools and designing theirs.
    }
    \item{
        It is equipped with Cryptography and security measures to ensure the confidentiality of patients' data in compliance with the latest standards.
    }
    \item{
        The machine learning toolkit abstraction also makes it easy to prepare on-demand analytical tools for the specific health-care application for which the dashboard is being designed.
    }
\end{itemize}

\section{Proposed Framework}
A remote health care application, in its core, requires advanced visualizations to be able to shed light on the underlying patterns of data, demonstrating temporal changes in the entity values, etc. It also demands the use of statistical analyses and data science approaches to help experts reach a better understanding of the statistical properties of the incoming data streams. At the same time, it needs to be able to receive information from the observer as well. In the following sections, the abstractions and their underlying mechanisms are elaborated upon.
These abstractions include visualization, data input, and machine learning capabilities, namely: Visualization Palette, Questionnaire Board, and Machine Learning Toolkit, respectively.

\subsection{Visualizations}
%todo
Perhaps the most significant component of remote monitoring applications is visualization, as it is often the main interface that assists the viewers in obtaining a thorough understanding of data from different aspects. In this work, the main visualization abstraction (namely, Visualization Palette) is created to help with the better encapsulation of this functionality from the rest of the RPM requirements.

This functionality presents itself as a dynamically updatable board along with an information form. The input form is designed in accordance with the specific needs of the application is data. Upon receiving the designated information set, the system will prepare the proper data stream for the web-application and ports it using the template engine to be shown on the client-side.

In this work, to better keep up with the most recent advancements in data visualization, we have chosen to implement the system using the latest version (v4.0) of the Data-Driven Documents javascript library \cite{bostock2011d3}. D3 allows the efficient implementation of well-formatted and interactive visualizations. At the same time, the system is also able to execute customized Python routine to generate and demonstrate the pictures on the web platform. This scheme allows developers to combine Python and D3.js library and covers the requirement for all possible visualizations to be implemented.

Depending on the application that the developers have in mind, any visualization with any properties can be implemented and added to the system by following a minimal series of steps. The idea is to provide a well-designed integration scheme in a way that requires the developers to focus merely on designing the components they are willing to add and not on integrating them into the system. The latter phase, namely, integration, will be automatically taken care of by the predefined functionalities designed for this matter. The descriptions and other information regarding the visualizations will also be added to the collection of visualizations that are displayed and discussed in the Portfolio panel, allowing the clients to be fully aware of the functionalities at hand and assisting them in utilizing them for their applications.

\subsection{Questionnaires}
The ability to inquire about the subjects and/or experts with regard to their observations is a critical aspect of many health monitoring applications \cite{mcgillicuddy2013patient}. With the advances in machine learning, these questionnaires and their collections of responses can be used as raw data that can be further processed and analyzed for ultimate inferences. However, in many applications, such direct observations and responses can be utilized as means for supervised or semi-supervised review of the other data sources in a fusion-based methodology. This means that these direct responses to the questions, which are often the resulting assessments of entities such as satisfaction or progress,  can be used as supervision signals to guide the learning processes better to traverse the optimization path in specific data processing approaches.

The ability to build questionnaires is considered a severe requirement in this work. Therefore, the questionnaires can be built using the abstraction provided under the title "Questionnaire Board." These questionnaires address different aspects of the monitoring platform, from receiving feedback on the data observed to gathering information and surveys. As another example use case, these questionnaires are often used for the developers to build an online web-application. Afterward, volunteers can utilize that app and tablets to ask people to fill with their information. This could range from psychotherapeutic tests to political opinion data.

\begin{figure}[h]
    \centering
    \includegraphics[width=0.49\textwidth]{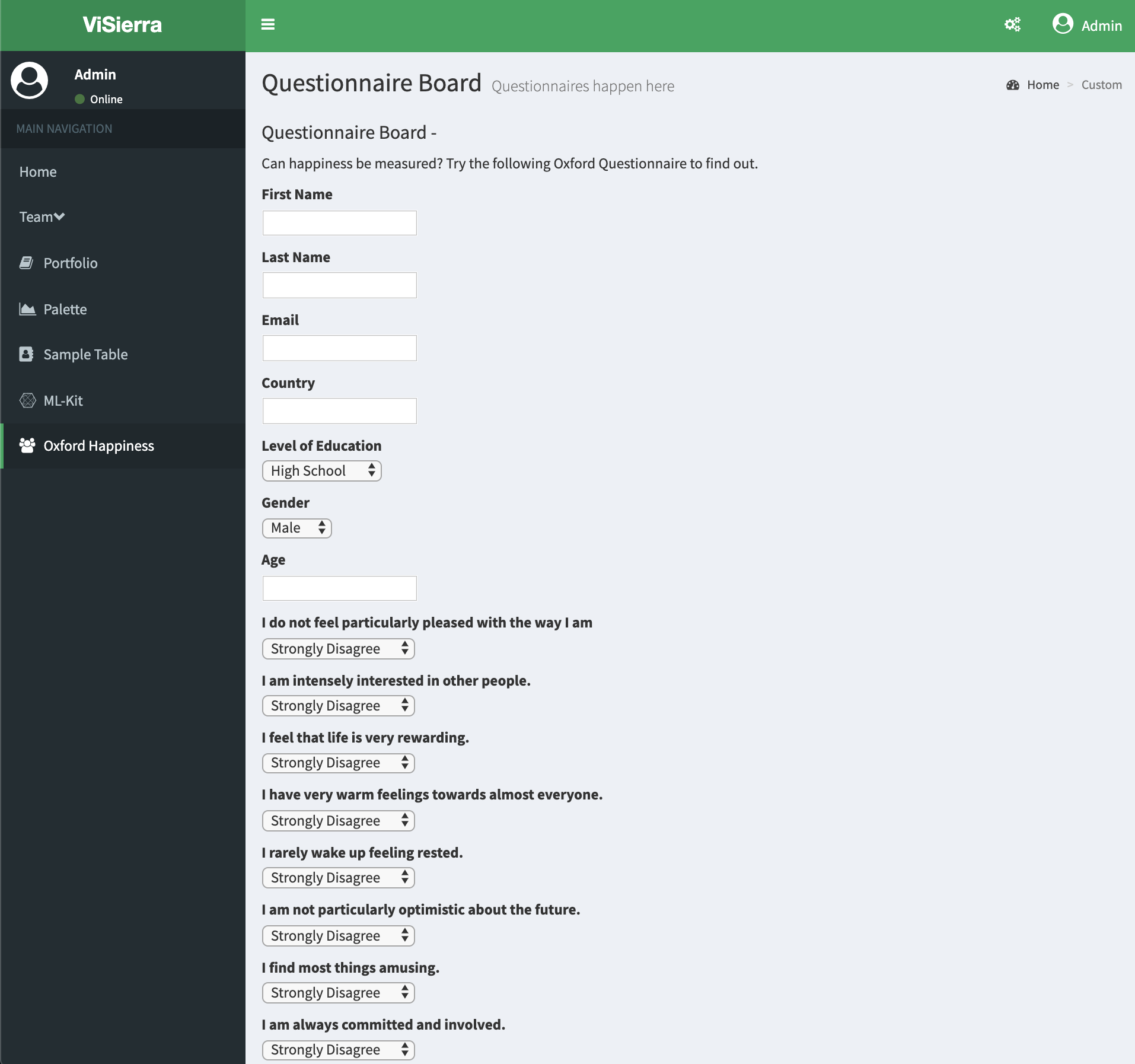}
    \caption{Sample ViSierra implementation of Oxford Happiness Questionnaire \cite{hills2002oxford}}
    \label{fig:my_label}
\end{figure}

\subsection{Machine Learning}
The ML Toolkit in this framework provides a place for the researchers to include the machine learning and data science components that are designed for the application. The general implementation includes fully connected neural networks that are shown to be an effective tool in building efficient predictors on medical data \cite{hosseini2018children}. The flask-based nature of the framework allows developers to integrate various pipelines from well-known deep learning libraries (PyTorch, TensorFlow, Sci-Kit Learn, etc.) into their system and maintain a channel to communicate information between their implemented ML pipelines and user feeds. Note that this framework utilizes several libraries that we have implemented earlier to help with providing developers with easy to use interfaces to more complicated analytics and dataframe manipulations \cite{plauthor,dataflame}.

\subsection{Other Functionalities}
Aside from the aforementioned components, which are the main parts of our proposed framework, other utilities (e.g., sheets for visualizing tables) are also implemented and can be used to expedite the research further and optimize the panels for better user experience.

\begin{figure}[h]
    \centering
    \includegraphics[width=0.49\textwidth]{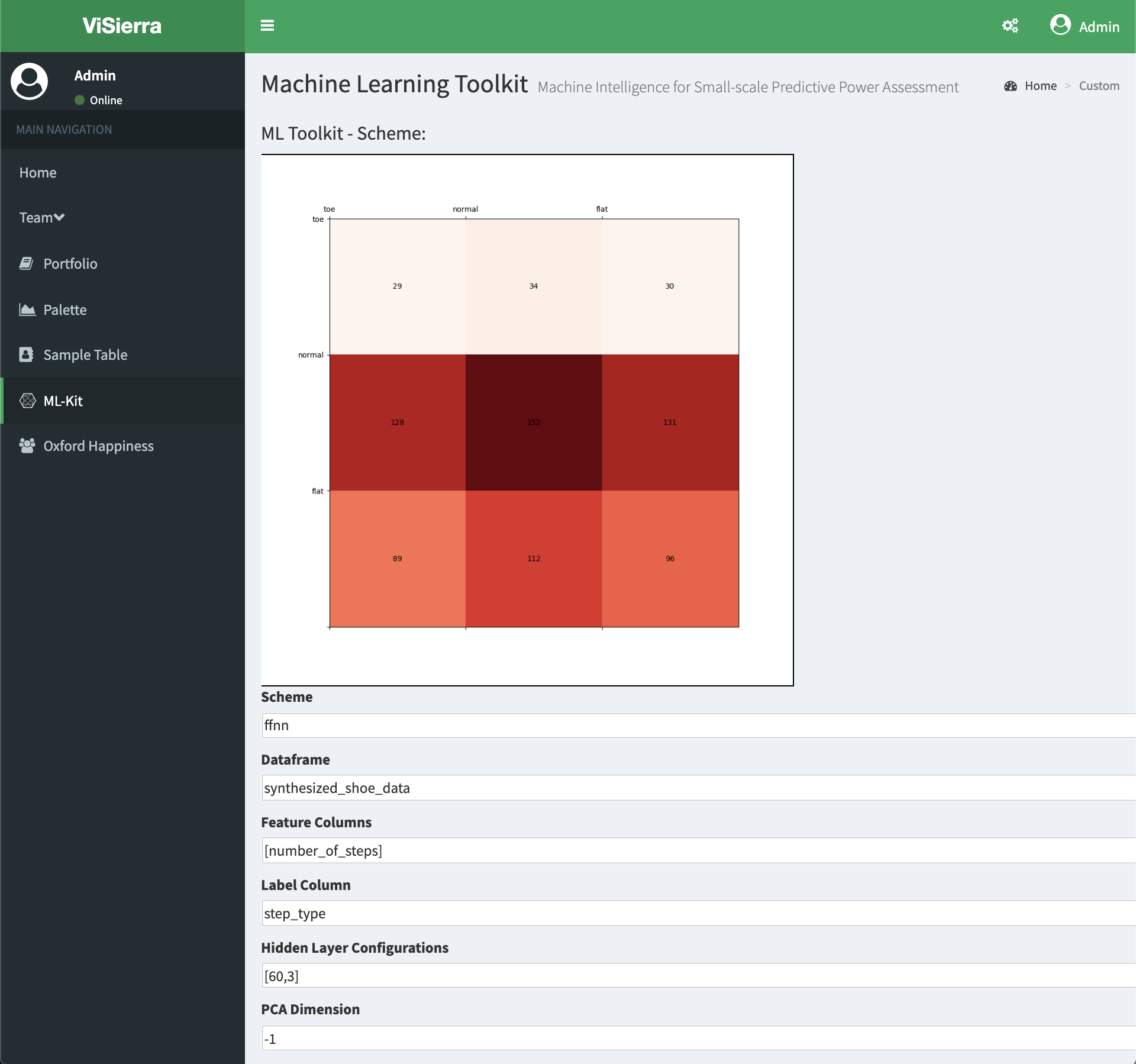}
    \caption{A sample usage of the machine learning toolkit - The output confusion matrix is visible in the figure, demonstrating the effectiveness in real applications}
    \label{fig:my_label}
\end{figure}

\section{Case Study: Exercise Tracking for Knee Rehabilitation}
The work on the remote monitoring of therapeutic knee exercises \cite{gwak2019extra} is an example of a real-world application in which such a framework presents useful. The objective in \cite{gwak2019extra} is to monitor the time-series of sensor values on knee braces to store the information on subjects' performance in therapeutic knee exercises.

An example use of this framework is in preparing the dashboard for this project. This dashboard can include the time-series visualizations for subjects on the selected dates and times, which can help the experts in determining the extent to which they were performed properly. It also can assist both subjects and experts in getting a better grasp of the improvements in the observed data.

\section{Discussion}
Remote health monitoring presents as the main component in almost every project in the thriving field of eHealth and wireless health. At its core, there are always dashboards to allow developers, researchers, and subjects to have access to the gathered data points and to allow them to input information as well. In this work, we presented ViSierra, a framework for RPM dashboards that provide a clean set of abstractions for different RPM components (visualization, questionnaires, machine learning algorithms, etc.), as well as offering a completely flexible environment based on Python and D3 library for the eHealth researchers to implement their dashboards in. A case study in which this framework can be utilized was also presented and discussed to demonstrate the effectiveness of this work.

% REFERENCES
\bibliography{./references.bib}{}
\bibliographystyle{IEEEtran}

\end{document}